\documentclass[12pt]{article}
\pdfoutput=1
\usepackage{url}
\usepackage{amssymb}
\usepackage{graphicx}
\usepackage{amsmath}
\begin{document}

\title{Haar Wavelets-Based Approach for Quantifying Credit Portfolio Losses}
\author{Josep J. Masdemont\thanks{Departament de Matem\`{a}tica Aplicada I, Universitat Polit\`{e}cnica de Catalunya,
        Diagonal 647, 08028 Barcelona, Spain. josep@barquins.upc.edu}
        \and Luis Ortiz-Gracia$^*$}
\date{April 2009}

\maketitle
\begin{abstract}
This paper proposes a new methodology to compute \emph{Value at Risk} (VaR) for quantifying losses in credit
portfolios. We approximate the cumulative distribution of the loss function by a finite combination of Haar
wavelets basis functions and calculate the coefficients of the approximation by inverting its Laplace transform.
In fact, we demonstrate that only a few coefficients of the approximation are needed, so VaR can be reached
quickly. To test the methodology we consider the Vasicek one-factor portfolio credit loss model as our model
framework. The Haar wavelets method is fast, accurate and robust to deal with small or concentrated portfolios,
when the hypothesis of the Basel II formulas are violated.
\end{abstract}

\renewcommand{\refname}{References}
\renewcommand{\figurename}{Figure}

\section{Introduction} It is very important for a bank to manage the risks originated from its business activities. In
particular, the credit risk underlying the credit portfolio is often the largest risk in a bank. The measured
credit risk is then used to assign capital to absorb potential losses arising from its
credit portfolio.\\

The Vasicek model is the basis of the Basel II IRB approach. It is a Gaussian one factor model such that default
events are driven by a latent common factor that is assumed to follow a Gaussian distribution, also called
\emph{Asymptotic Single Risk Factor} (ASRF) model. Under this model, loss only occurs when an obligor default in a
fixed time horizon. If we assume certain homogeneity conditions, this one factor model leads to a simple analytic
asymptotic approximation for the loss distribution and \emph{Value at Risk} (VaR). This approximation works well
for a large number of small exposures but can underestimate risks in the presence of
exposure concentrations.\\

Concentration risks in credit portfolios arise from an unequal distribution of loans to single borrowers
(\emph{name concentration}) or different industry or regional sectors (\emph{sector} or \emph{country
concentration}). Moreover, certain dependencies as, for example, direct business links between different
borrowers, can increase the credit risk in a portfolio since the default of one borrower can cause the default of
a dependent second borrower. This effect is called \emph{default contagion} and is linked to both name and sector
concentration.\\

In credit risk management one is particularly interested in the portfolio loss distribution. Since the portfolio
loss is usually modeled as a sum of random variables, the main task is to evaluate the probability density
function (PDF) of such a sum. The PDF of a sum of random variables is equal to the convolution of the respective
PDFs of the individual asset loss distributions. The evaluation of this convolution is a difficult problem
analytically, is computationally very intensive and in full generality is impractical for any
realistically sized portfolio.\\

For all these reasons, several methods have been developed in the last years. The saddle point approximation due
to \cite{Martin2001a} gives an analytical approximation of the Laplace inversion of the moment generating function
(MGF). This method has been improved by \cite{Martin2006} based on conditional independence models.
\cite{Glasserman2007} applies the methodology developed by \cite{Abate2000} to the single-factor Merton model.
First, the Bromwich integral is approximated by an infinite series using the trapezoidal rule and second, the
convergence of the infinite series is accelerated by a method called Euler summation. They have shown that the
cumulative distribution function (CDF) is comparatively accurate in small loss region, whereas the accuracy
worsens in the tail region. This is because the infinite series obtained by the Euler summation is an alternating
series, each
term of which has a very large absolute value.\\

Another approach to numerically invert the Laplace transform has been studied by \cite{Hoog1982} and
\cite{Ahn2003} consisting in applying the Poisson algorithm to approximate the Bromwich integral by an infinite
series, as in \cite{Abate2000} and then use the quotient-difference (QD) algorithm to accelerate the slow
convergence of the infinite series. We will refer to this approach as
\emph{the Hoog algorithm}. \cite{Takano2008}
has applied this methodology to the multi-factor Merton model. The numerical examples presented show that in
contrast with the Euler summation technique, de Hoog algorithm is quite efficient in measuring tail probabilities.\\

In this paper, we present a novel methodology for computing VaR through numerically inverting the Laplace
Transform of the CDF of the loss function once we have approximated it by a finite sum of Haar wavelets basis
functions. This kind of functions have compact support and so make them useful to study local properties of the
approximated function. Moreover, the CDF of
the loss function is discontinuous, making more suitable this way of approximation.\\

The remaining parts of the paper are organized as follows. In the next section we present the one-factor Gaussian
copula model and we define VaR as the risk measure used to quantify losses in Basel II Accord. In section three we
present the basic theory about Haar wavelets basis system used for the approximation detailed in section four.
Finally, we show with numerical examples the speed and accuracy of the new method in section five and section six
is devoted to conclusions.

\section{Portfolio Loss and Value at Risk} To represent the uncertainty about future events, we specify a
probability space $(\Omega,\mathcal{F},\mathbb{P})$ with sample space $\Omega$, $\sigma$-algebra $\mathcal{F}$,
probability measure $\mathbb{P}$ and with filtration $(\mathcal{F}_{t})_{t\geq0}$ satisfying the usual conditions.
We fix a time horizon $T>0$. Usually $T$ equals
one year.\\

Consider a credit portfolio consisting of $N$ obligors. Any obligor $n$ can be characterized by three parameters:
the \emph{exposure at default} $E_{n}$, the \emph{loss given default} which without loss of generality we assume
to be $100\%$ and the \emph{probability of default} $P_{n}$, assuming that each of them can be estimated from
empirical default data. The \emph{exposure at default} of an obligor denotes the portion of the exposure to the
obligor which is lost in case of default. Let $D_{n}$ be the default indicator of obligor $n$ taking the following
values

\begin{displaymath}
    D_{n} = \left\{\begin{array}{ll} 1, & \textrm{if obligor $n$ is
    in default,}\\
    0, & \textrm{if obligor $n$ is not in default,}\end{array} \right.
\end{displaymath}

Let $L$ be the portfolio loss given by:
$$ L = \sum_{n=1}^{N} L_{n}, $$
where $L_{n} = E_{n} \cdot D_{n}$.\\

To test our methodology we consider the Vasicek one-factor Gaussian copula model as our model framework. The
Vasicek model is a one period default model, i.e., loss only occurs when an obligor defaults in a fixed time
horizon. Based on Merton`s firm-value model, to describe the obligor's default and its correlation structure, we
assign each obligor a random variable called firm-value. The firm-value of obligor $n$ is represented by a common,
standard normally distributed factor $Y$ component and an idiosyncratic standard normal noise component
$\epsilon_{n}$. The $Y$ factor is the state of the world or business cycle, usually called systematic factor.

\begin{equation*}
    V_{n}(T)= \sqrt{\rho_{n}} Y+\sqrt{1-\rho_{n}} \epsilon_{n},
\end{equation*}
where $Y$ and $\epsilon_{n}$, $\forall n\le{N}$ are i.i.d. standard
normally distributed.\\

In case that $\rho_{n}=\rho$ for all $n$, the parameter $\rho$ is called the common asset correlation. The
important point is that conditional on the realization of the systematic factor $Y$, the
firm's values and the defaults are independent. From now on, we assume $\rho_{n}$ to be constant.\\

Let us explain in detail the meaning of systematic and idiosyncratic risk.  The first one can be viewed as the
macro-economic conditions and affect the credit-worthiness of all obligors simultaneously.  The second one represent
conditions inherent to each obligor and this is why they are assumed to be
independent of each other.\\

In the Merton model, each obligor $n$ defaults if its firm-value falls below the threshold level $T_{n}$ defined
by $T_{n} \equiv {\Phi^{-1}(P_{n})}$ where $\Phi(x)$ is the standard normal cumulative distribution function and
$\Phi^{-1}(x)$ denotes its inverse function. The probability of obligor $n$'s default conditional on a realization
of $Y=y$ is given by

\begin{equation*}
    p_{n}(y)\equiv{\mathbb{P}(V_{n}<T_{n}\mid Y=y)}=\Phi\left(\frac{T_{n}-\sqrt{\rho}y}{\sqrt{1-\rho}}\right).
\end{equation*}

Consequently, the conditional probability of default depends on the systematic factor, reflecting the fact that
the business cycle
affect the possibility of an obligor's default.\\

Let $F$ be the cumulative distribution function of $L$. Without loss of generality, we can assume $\sum_{n=1}^{N}
E_{n} = 1$ and consider

\begin{displaymath}
    F(x) = \left\{\begin{array}{ll} \overline{F}(x), & \textrm{if $0\leqslant x \leqslant 1$,}\\
    1, & \textrm{if $x>1$,}\end{array} \right.
\end{displaymath}
for a certain $\overline F$ defined in $[0,1]$.\\

Let $\alpha \in (0,1)$ be a given confidence level, the $\alpha$-quantile of the loss distribution of $L$ in this
context is called \emph{Value at Risk} (VaR). Thus,
$$ l_{\alpha}=inf\{l \in \mathbb{R}: \mathbb{P}(L \le l) \ge \alpha \} = inf\{l \in \mathbb{R}: F(l)\ge \alpha \}. $$
Usually the $\alpha$ of interest is very close to 1. This is the measure chosen in the Basel II Accord for the
computation of capital requirement, which means a bank that manages its risks with Basel II must to reserve
capital by an amount of
$l_{\alpha}$ to cover extreme losses.\\

\section{The Haar Basis Wavelets System} Consider the space $L^{2}(\mathbb{R})=\{ f: \int_{-\infty}^{+\infty}\left|f(x)\right|^2dx<\infty
\}$. For simplicity, we can view this set as a set of functions $f(x)$ which
get small in magnitude fast enough as $x$ goes to plus and minus infinity.\\

A general structure for wavelets in $L^{2}(\mathbb{R})$ is called a \emph{Multi-resolution Analysis} (MRA). We
start with a family of closed nested subspaces
\[ ... \subset V_{-2} \subset V_{-1} \subset V_{0} \subset V_{1} \subset V_{2} \subset ... \]
in $L^{2}(\mathbb{R})$ where
\[ \displaystyle\bigcap_{j\in\mathbb{Z}}{V_{j}}=\{0\}, \hspace{1cm} \displaystyle\overline{\bigcup_{j\in\mathbb{Z}}{V_{j}}}=L^{2}(\mathbb{R}), \]
and
\[ f(x) \in V_{j} \Longleftrightarrow f(2x) \in V_{j+1}. \]
If these conditions are met, then there exists a function $\phi \in V_{0}$ such that
$\{\phi_{j,k}\}_{k\in\mathbb{Z}}$ is an orthonormal basis of $V_{j}$, where
\[ \phi_{j,k}(x)=2^{j/2}\phi(2^{j}x-k). \]
In other words, the function $\phi$, called the \emph{father function}, will generate an orthonormal basis for
each $V_{j}$ subspace.\\

Then we define $W_{j}$ such that $V_{j+1}=V_{j} \oplus W_{j}$. This says that $W_{j}$ is the space of functions in
$V_{j+1}$ but not in $V_{j}$, and so, $L^{2}(\mathbb{R})=\sum_{j} \oplus W_{j}$. Then (see \cite{Daubechies1992})
there exists a function $\psi \in W_{0}$ such that $\{\psi_{j,k}\}_{k\in\mathbb{Z}}$ is an orthonormal basis of
$W_{j}$, and $\{\psi_{j,k}\}_{j,k\in\mathbb{Z}}$ is a wavelets basis of $L^{2}(\mathbb{R})$, where
\[ \psi_{j,k}(x)=2^{j/2}\psi(2^{j}x-k). \]
The function $\psi$ is called the \emph{mother function}, and the $\psi_{j,k}$ are the \emph{wavelets functions}.\\

For any function $f \in L^{2}(\mathbb{R})$ a projection map of $L^{2}(\mathbb{R})$ onto $V_{m}$,
\[ \mathcal{P}_{m}:L^{2}(\mathbb{R}) \rightarrow V_{m}, \] is defined by

\begin{equation}\label{app_def}
 \mathcal{P}_{m}f(x)  =\sum_{j=-\infty}^{m-1} \sum_{k=-\infty}^{k=+\infty} d_{j,k}\psi_{j,k}(x) \\
        =\sum_{k\in\mathbb{Z}}c_{m,k}\phi_{m,k}(x).
\end{equation}
where $d_{j,k}=\int_{-\infty}^{+\infty}f(x)\psi_{j,k}(x)dx$ are the wavelets coefficients and the
$c_{m,k}=\int_{-\infty}^{+\infty}f(x)\phi_{m,k}(x)dx$ are the scaling coefficients. The first part in
(\ref{app_def}) is a truncated wavelets series. If $j$ were allowed to go to infinity, we would have the full
wavelets summation. The second part in (\ref{app_def}) gives an equivalent sum in terms of the scaling functions
$\phi_{m,k}$. Considering higher $m$ values, meaning that more terms are used,
the truncated series representation of our function improves.\\

To develop our work, we have used Haar wavelets  (see \cite{Daubechies1992}). For these wavelets, the space
$V_{j}$ is the set of all $L^{2}(\mathbb{R})$ functions which are constant on each interval of the form
$[\frac{k}{2^j},\frac{k+1}{2^j})$ for all integers $k$. Then

\begin{displaymath}
    \phi(x) = \left\{\begin{array}{ll} 1, & \textrm{if $0 \leq x < 1$,}\\
    0, & \textrm{otherwise,}\end{array} \right.
\end{displaymath}

and

\begin{displaymath}
    \psi(x) = \left\{\begin{array}{ll} 1, & \textrm{if $0 \leq x < \frac{1}{2}$,}\\
              -1, & \textrm{if $\frac{1}{2} \leq x < 1$,}\\
              0, & \textrm{otherwise.}\end{array} \right.
\end{displaymath}
The unique thing about using wavelets as opposed to Fourier series is that the wavelets can be moved (by the $k$
value), stretched or compressed (by the $j$ value) to accurately represent a function local properties. Moreover,
$\phi_{j,k}$ is nonzero only inside the interval $[\frac{k}{2^j},\frac{k+1}{2^j})$. These facts will be used later
to compute the VaR without the need of knowing the whole distribution of the loss function.

\begin{figure}
\begin{center}
\begin{tabular}{cc}
\includegraphics[height=6cm,width=6.5cm]{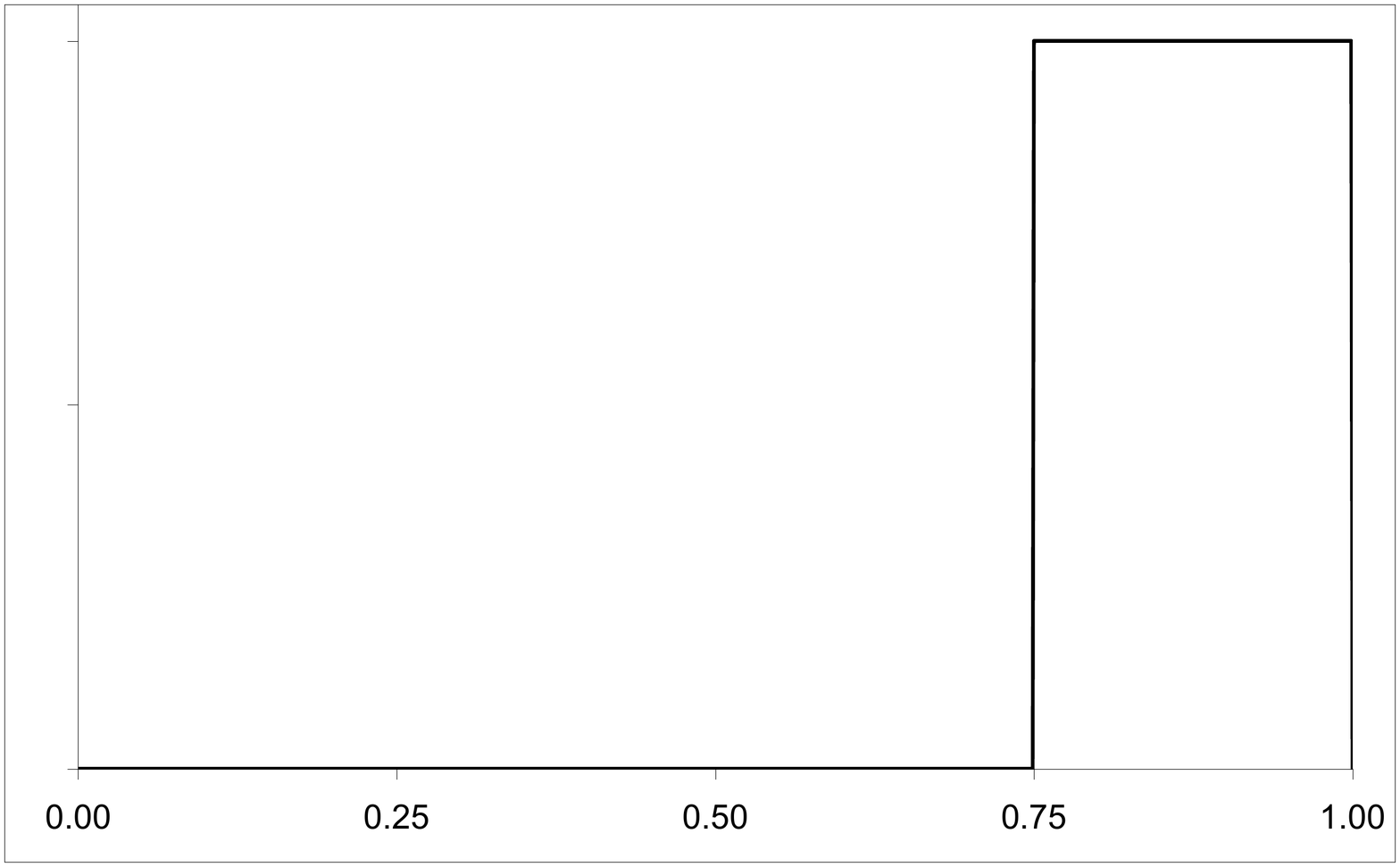} &
\includegraphics[height=6cm,width=6.5cm]{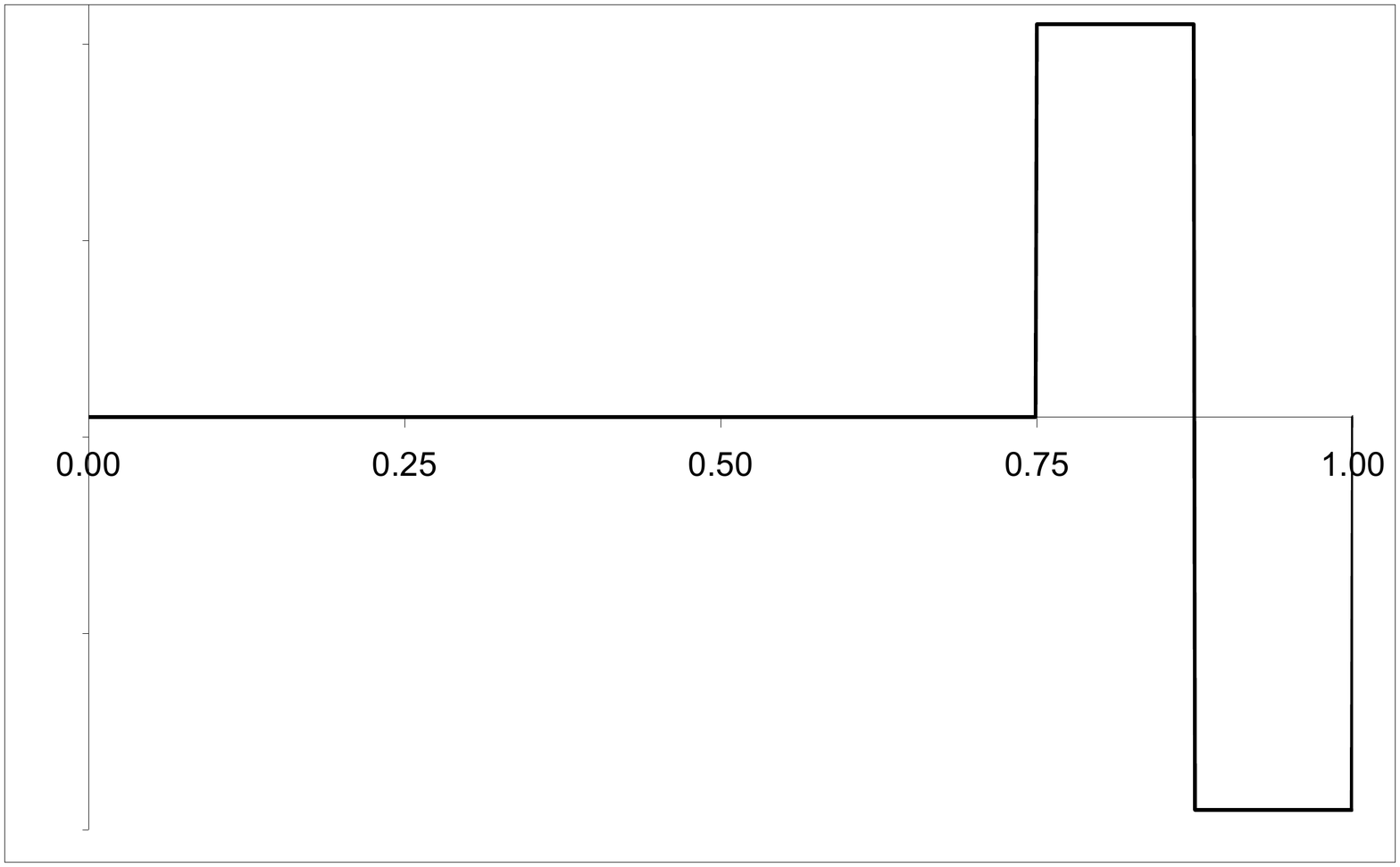}
\end{tabular}
\end{center}
\caption{Scaling ($\phi_{2,3}$) and wavelet ($\psi_{2,3}$) functions.} \label{fig-wav}
\end{figure}

\section{Haar wavelets approximation} Let us mention an issue regarding the CDF $\overline F$ defined
above. Since the loss can take only a finite number of discrete values ($2^N$ at most) the PDF of the loss
function is a sum of Dirac delta functions and then, the CDF is a discontinuous function. Moreover, the stepped
form of the CDF makes the Haar wavelets a natural and very well-suited way
of approximation.

\subsection{Laplace Transform Inversion} Due to the fact that $\overline F \in L^{2}([0,1])$ and according to the theory of MRA,
we can approximate $\overline F$ in $[0,1]$ by a sum of scaling functions,

\begin{equation}
 \overline F(x) \approx \sum_{k=0}^{2^m-1} c_{m,k}\phi_{m,k}(x), \label{approx_F}
\end{equation}

and

\[ \overline F(x) = \lim_{m\rightarrow+\infty} \sum_{k=0}^{2^m-1} c_{m,k}\phi_{m,k}(x). \]

Recall that in the one-factor model framework, if the systematic factor $Y$ is fixed, default occurs independently
because the only remaining uncertainty is the idiosyncratic risk. The MGF conditional on $Y$ is thus given by the
product of each obligor's MGF as

\begin{equation*}
\begin{split}
M_{L}(s;Y) \equiv \mathbb{E}(e^{-sL} \mid Y) & = \prod_{n=1}^{N}\mathbb{E} \left (e^{-sE_{n}D_{n}} \mid Y \right) \\
                                            & = \prod_{n=1}^{N} \left [ 1-p_{n}(y)+p_{n}(y)e^{-sE_{n}} \right ].
\end{split}
\end{equation*}

Notice that we are assuming non stochastic LGD. Taking the expectation value
of this conditional MGF yields the unconditional MGF,

\begin{equation*} \label{y_integral}
\begin{split}
M_{L}(s) \equiv \mathbb{E}(e^{-sL})& = \mathbb{E}(\mathbb{E}(e^{-sL} \mid Y)) \\
                               & = \mathbb{E}(M_{L}(s;Y))=\mathbb{E}
\left [ \prod_{n=1}^{N} \left [ 1-p_{n}(y)+p_{n}(y)e^{-sE_{n}} \right ] \right ] \\ & = \int_{\mathbb{R}}
\prod_{n=1}^{N} \left [ 1-p_{n}(y)+p_{n}(y)e^{-sE_{n}} \right ]\frac{1}{\sqrt{2\pi}}e^{-\frac{y^2}{2}}dy.
\end{split}
\end{equation*}

But if $f$ is the probability density function of the loss function then the unconditional MGF is also the Laplace
transform of $f$:

\begin{equation}
M_{L}(s) \equiv \mathbb{E}(e^{-sL}) = \int_{0}^{+\infty} e^{-sx}f(x)dx =
         \widetilde{f}(s). \label{tl_f}
\end{equation}

As we have noticed before,

\begin{equation} \label{deltas}
 f(x)=\sum_{i=0}^{2^N}\mu_{i}\delta(x-x_{i}), \hspace{1cm} x_{1},x_{2},...,x_{2^N} \in [0,1].
\end{equation}
where $\delta(x-x_{i})$ is the Dirac delta at $x_i$ that can be thought as a density distribution of a unit of
mass concentrated in the point $x_i$ (i.e. $\int_{0}^{+\infty} g(x) \delta(x-x_{i}) dx=g(x_i)$, for every test
function $g(x)$). Probabilistically, a distribution, such as (\ref{deltas}), corresponds to a situation where only
the scenarios $x_{1},x_{2},...,x_{2^N}$ are feasible with respective probabilities
$\mu_{1},\mu_{2},...,\mu_{2^N}$. Of course these probabilities must be positive and sum up 1, this is,
$$ \sum_{i=0}^{2^N}\mu_{i}=1. $$

As it is also well known, in the context of generalized functions, the derivative
of the Heaviside step function is a Dirac delta. In this context (and of course
in the context of regular functions) we can integrate by parts the expression
(\ref{tl_f}) and using the approximation (\ref{approx_F}) to conclude that,
\begin{equation}
\begin{split}
M_{L}(s) & = \int_{0}^{+\infty}e^{-sx}F'(x)dx = e^{-s}+s\int_{0}^{1}e^{-sx}\overline F(x)dx \\
         & \approx e^{-s}+s\int_{0}^{1}\left[e^{-sx}\sum_{k=0}^{2^m-1} c_{m,k}\phi_{m,k}(x) \right]dx\\
         & = e^{-s} + 2^{\frac{m}{2}}s\sum_{k=0}^{2^m-1}c_{m,k}\widetilde{\phi}_{m,k}(s), \label{mgf1}
\end{split}
\end{equation}

where
\[\widetilde{\phi}_{m,k}(s)=\frac{1}{s}e^{-s\frac{k}{2^m}}(1-e^{-s\frac{1}{2^m}}).\]
is the Laplace transform of the basis function $\phi_{m,k}(x)$.\\

Observing that $\widetilde{\phi}_{m,k}(s)=\widetilde{\phi}_{m,0}(s)e^{-s\frac{k}{2^m}}$ and making the change of
variable $z=e^{-s\frac{1}{2^m}}$, the expression (\ref{mgf1}) is the same as

\begin{equation} \label{def_Q}
Q(z)\equiv \sum_{k=0}^{2^m-1}c_{m,k}z^k \approx \frac{M_{L}(-2^mln(z))-z^{2^m}}{2^{\frac{m}{2}}(1-z)}.
\end{equation}
Where we note that for $r<1$, $Q(z)$ is analytic inside the disc of the complex
plane $\{z:\left|z\right|<r\}$, since the singularity in $z=0$ is avoidable.
Then, given the generating function $Q(z)$, we can obtain expressions for the
coefficients $c_{m,k}$ by means of the Cauchy's integral formula. This is,

\[ c_{m,k}=\frac{1}{2\pi i}\int_{\gamma}\frac{Q(z)}{z^{k+1}}dz, \hspace{1cm} k=0,1,...,2^{m}-1. \]
where $\gamma$ is a circle about the origin of radius $r$, $0<r<1$.\\

Making the change of variable $z=re^{iu}$, $0<r<1$,

\begin{equation} \label{trapecios}
\begin{split}
c_{m,k}= \frac{1}{2\pi r^k}\int_{0}^{2\pi}\frac{Q(re^{iu})}{e^{iku}}du
       & = \frac{1}{2\pi r^k}\int_{0}^{2\pi}\left[ Re(Q(re^{iu}))cos(ku)+Im(Q(re^{iu}))sin(ku) \right]du \\
       & =\frac{2}{\pi r^k}\int_{0}^{\pi}Re(Q(re^{iu}))cos(ku)du.
\end{split}
\end{equation}

Finally, we can calculate the integral in (\ref{trapecios}) approximately by
means of the trapezoidal rule to obtain the coefficients.\\

\subsection{VaR computation}
It can be easily proved that \[ 0\leq c_{m,k} \leq 2^{-\frac{m}{2}}, \hspace{1cm} k=0,1,...,2^{m}-1, \] and \[ 0
\leq c_{m,0} \leq c_{m,1} \leq ...\leq c_{m,2^m-1}.\] \\

VaR can now be calculated quickly with at most $m$ coefficients for each fixed level of resolution $m$ of the
approximation, due to the compact support of the basis functions. Observe that

\[ \overline F(l_{\alpha})=2^{\frac{m}{2}} \cdot c_{m,k} \hspace{1cm} \frac{k}{2^m} \leq l_{\alpha} \leq \frac{k+1}{2^m} \]

Thus, we can simply start searching $l_{\alpha}$ computing $\overline F(\frac{2^{m-1}}{2^{m}})$ with any bisection like method.
If $\overline F(\frac{2^{m-1}}{2^{m}})>\alpha$ then we compute
$\overline F(\frac{2^{m-1}-2^{m-2}}{2^{m}})$, otherwise we compute
$\overline F(\frac{2^{m-1}+2^{m-2}}{2^{m}})$, and so on. Finally, observe that VaR lies between $\frac{k}{2^m}$
and $\frac{k+1}{2^m}$ for a certain $k \in \{0,1,...,2^{m-1}\}$.

\section{Numerical Examples}
In this section we present a comparative study to calculate VaR between the
Wavelet Approximation (WA) method and Monte Carlo (MC).
As it is well known, MC has a strong dependence between the size of the
portfolio and the computational time. As the size increases, MC becomes a
big time consuming method.\\

The real situation in some financial companies show us that there are strong concentrations in their credit
portfolios. Basel II formulas to calculate VaR are supported under unrealistic hypothesis, such as infinite number
of obligors with small exposures. \\

For these reasons, we test our methodology with concentrated portfolios. We consider four portfolios ranging from
100 to 10000 obligors with the main numerical results displayed in Table~\ref{tabla}. The Wavelet Approximation
with $m=10$ provides accurate results in a few seconds of computational time\footnote{Computations have been
implemented in C under a PC with Intel CPU 280 GHz and 496 MB RAM.}, since the 
maximum relative error $(\max \left|\hbox{WA}\hbox{WA}\right|/MC)$
in portfolios P1, P2 and P3 is only 0.5\%, 1\% and 0.7\% respectively. The 
relative error in portfolio P4 may look somewhat greater than expected (2.5\%).
But in this case, performing $5\times 10^5$ MC simulations we find that the
VaR obtained is 0.160, being the maximum relative error of 1.2\% and showing 
again the fast convergence of the WA methodology.\\

\begin{figure}[t!]
\begin{center}
\includegraphics[width=130mm]{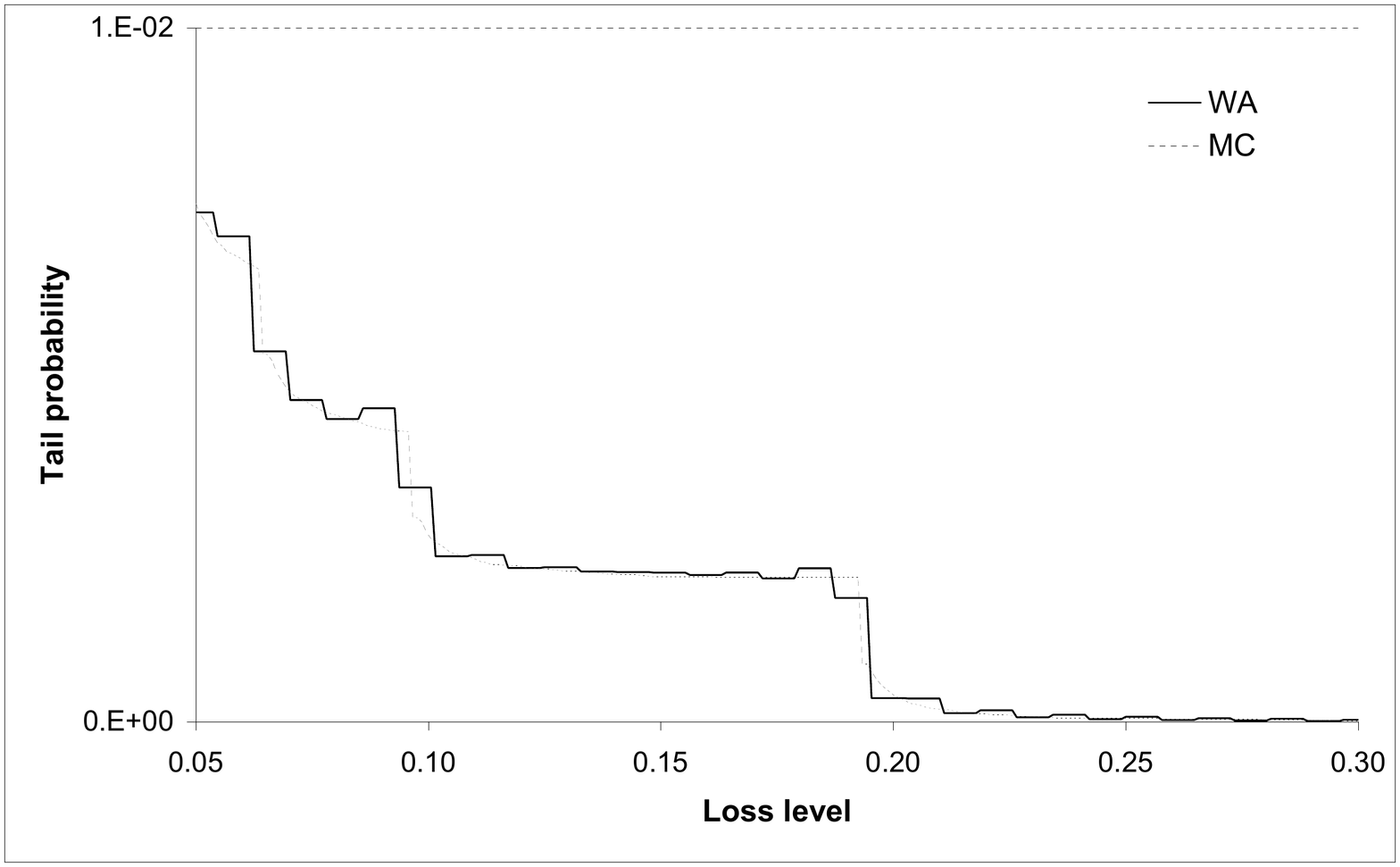}
\end{center}
\caption{Tail probability approximation of portfolio P1 with $m=7$.} \label{fig-1}
\end{figure}

\begin{figure}[t!]
\begin{center}
\includegraphics[width=130mm]{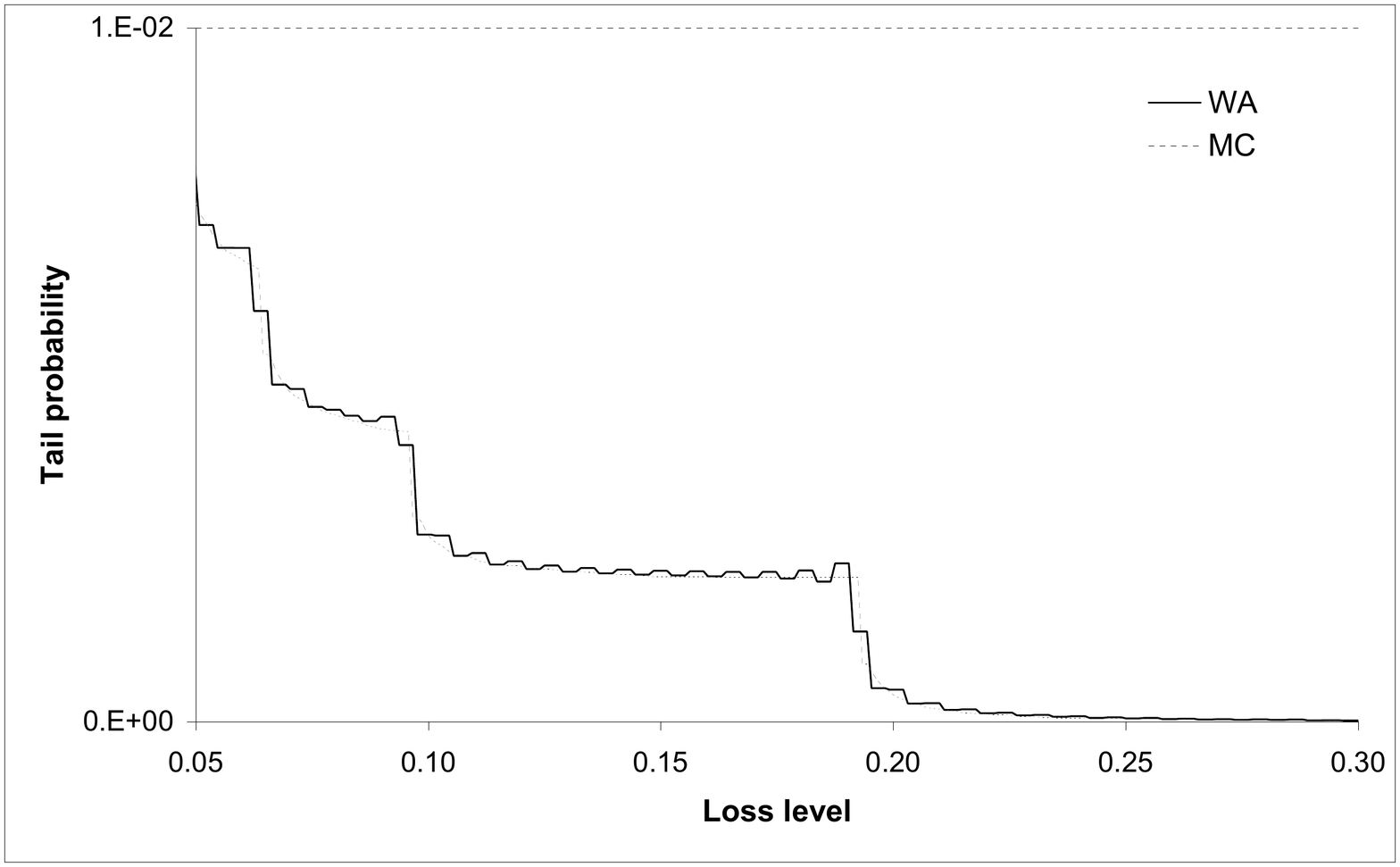}
\end{center}
\caption{Tail probability approximation of portfolio P1 with $m=8$.} \label{fig-2}
\end{figure}

\begin{figure}[t!]
\begin{center}
\includegraphics[width=130mm]{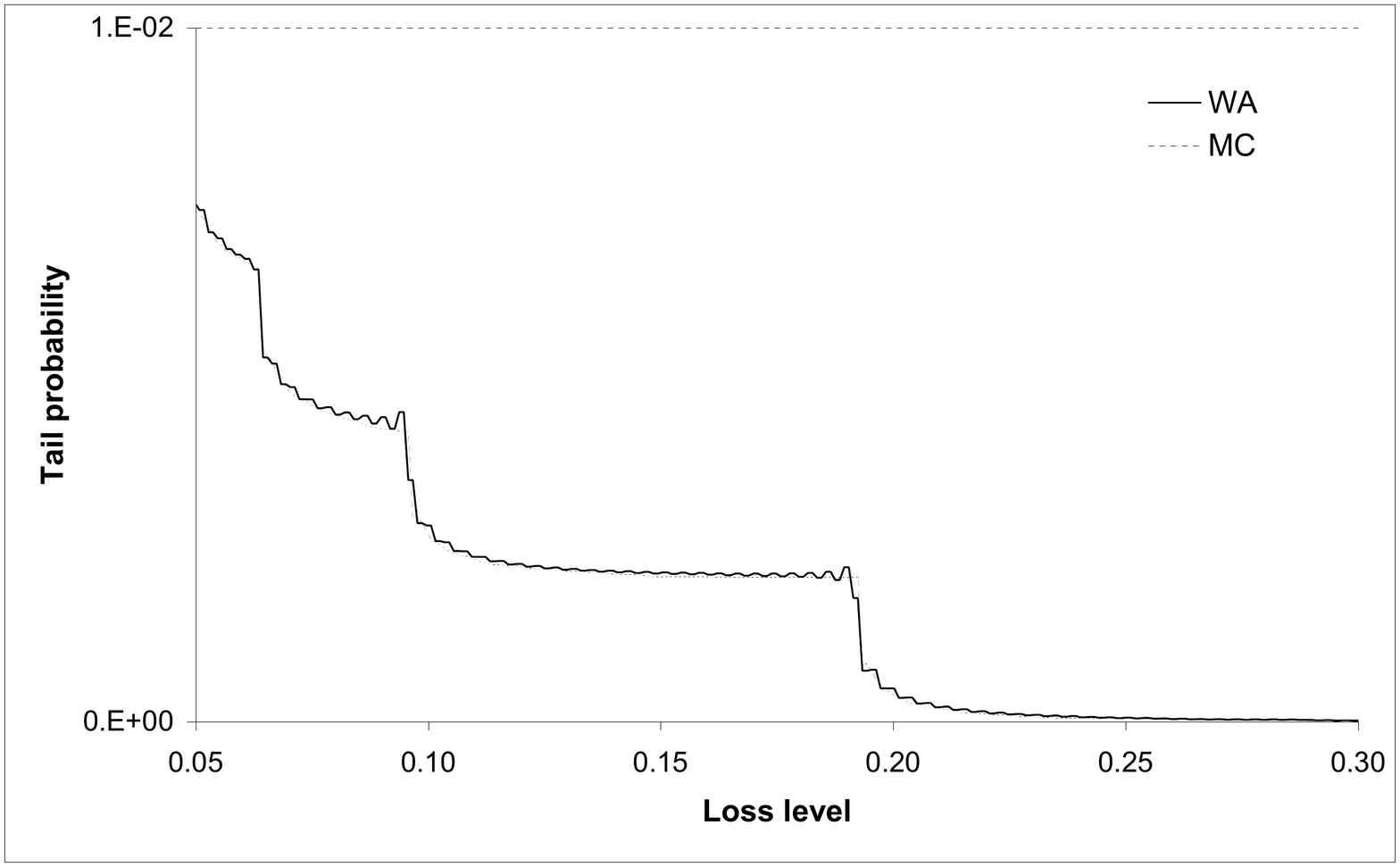}
\end{center}
\caption{Tail probability approximation of portfolio P1 with $m=9$.} \label{fig-3}
\end{figure}

\begin{figure}[t!]
\begin{center}
\includegraphics[width=130mm]{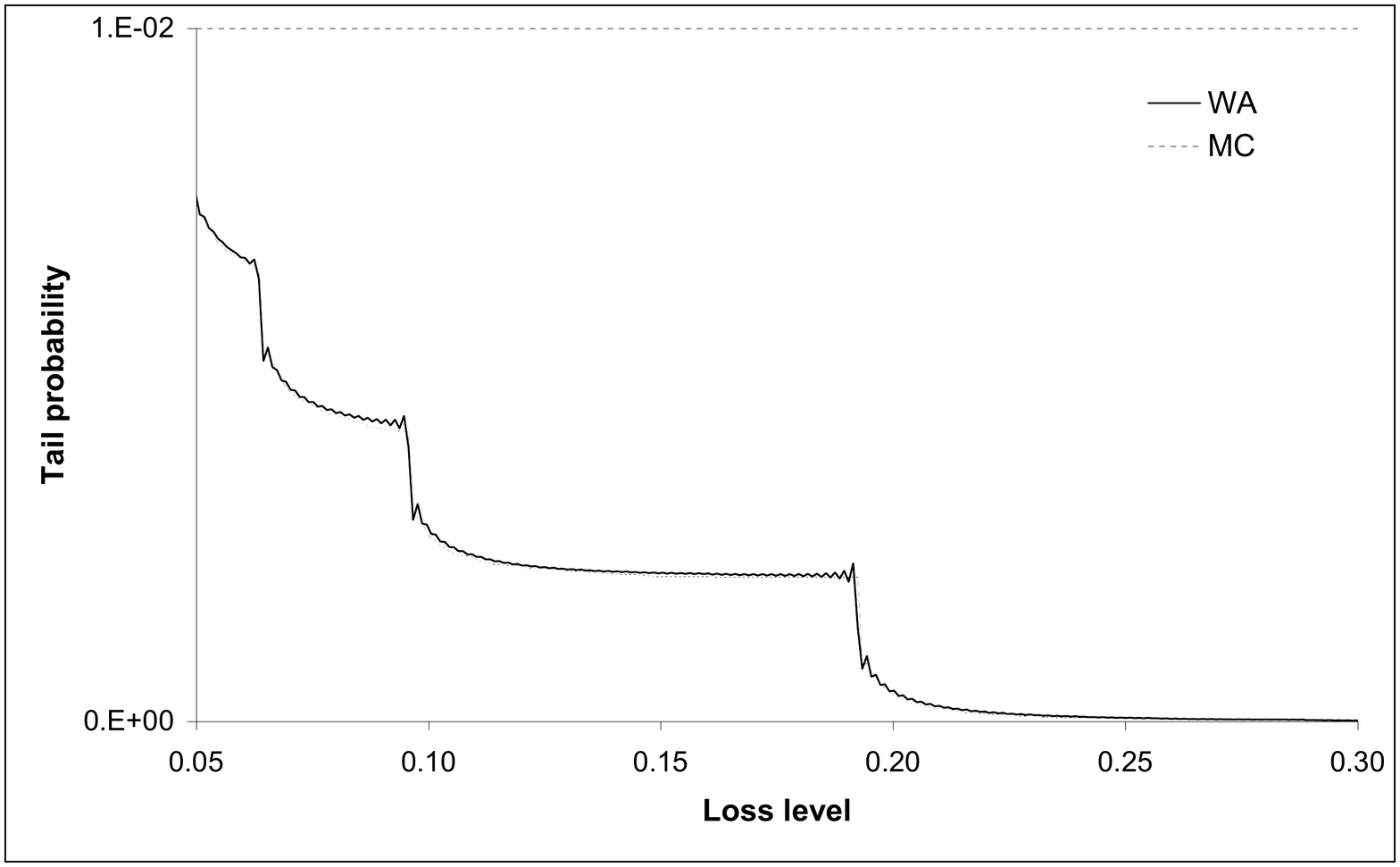}
\end{center}
\caption{Tail probability approximation of portfolio P1 with $m=10$.} \label{fig-4}
\end{figure}

Finally, in order to display the accuracy of Wavelet Approximation, we have considered the tail probabilities of
the loss function associated with portfolio P1 using different resolutions ($m=7,8,9$ and 10). The results,
compared again with Monte Carlo simulations, can be seen in figures~\ref{fig-1},\ref{fig-2},\ref{fig-3} and
\ref{fig-4} respectively. It is remarkable how the Haar wavelets are naturally
capable of detecting jumps in the cumulative
distribution, making the approximation very precise with not many terms.

\begin{table}
\begin{center}
\begin{tabular}{c c c c c c}
\hline\hline Portfolio & N & $P_{n}$ & WA $(m=9)$ & WA $(m=10)$ & MC \\
\hline\hline
P1 & 100 & 0.21\% & $[0.193,0.195)$ & $[0.194,0.195)$ & 0.194 \\
[1ex] P2 & 1000 & 1.00\% & $[0.191,0.193)$ & $[0.192,0.193)$ & 0.194 \\ [1ex] P3 & 1000 & 0.30\% & $[0.139,0.141)$
& $[0.140,0.141)$ & 0.141 \\ [1ex] P4 & 10000 & 1.00\% & $[0.160,0.162)$ & $[0.161,0.162)$ & 0.158
\\ [1ex]\hline
\end{tabular}
\end{center}
\caption{Results of 99.9\% VaR computation with Wavelet Approximation and Monte Carlo simulations with $2\times
10^5$ random scenarios and $\rho=0.15$. In order to consider concentrated
portfolios, in all cases we have taken $E_{n}=\frac{C}{n}$ where $C$ is a 
constant such that $\sum_{n=1}^{N}E_{n}=1$.}\label{tabla} \centering
\end{table}

\section{Conclusions}
We have presented a numerical approximation to the loss function based on Haar wavelets system. First of all, we
approximate the discontinuous distribution of the loss function by a finite sum of Haar scaling functions, and
then we calculate the coefficients of the approximation by inverting its Laplace transform. Due to the compact
support property of Haar system, only a few coefficients are needed for VaR computation.\\

We have shown the performance of the numerical approximation in four sample
portfolios. These results among other simulations, show that the
method is applicable to different sized portfolios and very accurate
with short time computations. Moreover, the Wavelet Approximation is robust
since the method is very stable under changes in the parameters of the model.
The stepped form of the approximated distribution makes the Haar wavelets
natural and very suitable for the approximation.\\

We also remark that the algorithm is valid for continuous cumulative
distribution functions, and that it can be used in other financial models
without making conceptual changes in the development. For instance, we can
easily introduce
stochastic loss given default (just changing a bit the unconditional moment
generating function) and consider the multi-factor Merton model as the model
framework as well.

\end{document}